\documentclass[preprint2]{aastex}

\usepackage{lscape}
\usepackage{apjfonts}
\begin{document}


\title{First Lunar Occultation
Results from the 2.4\,m Thai National Telescope equipped with
ULTRASPEC
}


\author{A. Richichi\altaffilmark{1}, 
P. Irawati\altaffilmark{1}, 
B. Soonthornthum\altaffilmark{1}, 
V.S. Dhillon\altaffilmark{2}, 
T.R. Marsh\altaffilmark{3} 
}


\altaffiltext{1}{National Astronomical Research Institute of Thailand, 191 Siriphanich Bldg., 
Huay Kaew Rd., Suthep, Muang, Chiang Mai  50200, Thailand}
\altaffiltext{2}{Department of Physics and Astronomy, University of Sheffield, 
Sheffield S3 7RH, UK}
\altaffiltext{3}{Department of Physics, University of Warwick, 
Gibbet Hill Road, Coventry CV4 7AL, UK}

\email{andrea4work@gmail.com}


\begin{abstract}
The recently inaugurated 2.4\,m Thai National Telescope (TNT) is
equipped, among other instruments, with the
ULTRASPEC low-noise, frame-transfer EMCCD camera.
At the end of its first official observing season, we report
on the use of this facility to record high time resolution
imaging using small detector subarrays with sampling as
fast as several $10^2$\,Hz.
In particular, we have recorded lunar occultations of
several stars which represent the first contribution to this
area of research made
from South-East Asia with a telescope of this class.
Among the results, we discuss an accurate measurement of
$\alpha $~Cnc, which has been reported previously as a
suspected close binary. Attempts to resolve this star by
several authors have so far met with a lack of
unambiguous confirmation. With our observation 
we are able to place stringent limits
on the projected angular separation ($<0\farcs003$) and brightness  ($\Delta{\rm m} > 5$) of a 
putative companion.
We also present a measurement of the binary
{HR~7072}, which extends considerably the time
coverage available for its yet undetermined orbit.
We discuss our precise determination of the flux ratio
and projected separation in the context of other
available data.
We conclude by providing an estimate of
the performance of ULTRASPEC at TNT for lunar occultation
work. This facility can
help to extend the lunar occultation technique in a geographical area where
no comparable resources were available until now.
\end{abstract}





\section{Introduction}
Lunar occultations (LO) are 
a method
to obtain high angular resolution by means of the
analysis of the diffraction pattern generated when a background
source is occulted by the lunar limb. One major advantage of
LO is the fact that the angular resolution
achieved is mainly dependent on relations between the
wavelength and the distance to the Moon, rather than
the aperture of the telescope as in standard imaging. Thus, the
measurement of angular sizes well beyond the diffraction
limit of even the largest telescopes on the ground or
in space becomes possible. Significant limitations of the
method are that the sources cannot be chosen at
will, that the events are time-critical, and that opportunities
to repeat the observations are rare.
In order to detect and measure the diffraction fringes,
it is necessary to sample the light curves with
time resolutions of order 1\,ms.

At least three major goals can
be achieved by LO: the measurement of stellar
angular diameters, the detection and characterization
of circumstellar envelopes and extended emissions, 
and finally the study of small-separation binary stars. 
For decades, LO have made 
contributions in the first two areas above,
but
currently  long-baseline
interferometry, more time consuming but more complete
in its results, is often used  \citep[see CHARM2 catalog,][]{CHARM2}. 
Concerning
binary stars, however, LO maintain an edge in that they
are quick to observe, easy to analyze, and offer
a combination of sensitivity and dynamic range which
remains unsurpassed.  
In the
course of a long program of occultations with the ISAAC
instrument at the ESO VLT,
\citet[and references therein]{2014AJ....147...57R} have recorded
over a thousand occultation
events with an $\approx 8$\% detection rate of mostly new
binary or multiple systems, the majority of which 
are out of reach
of other techniques. The ISAAC instrument has now been
decommissioned, and only few observatories 
around the world remain  suitably
equipped to observe LO.

In this paper we first introduce the Thai National
Telescope (TNT), in conjunction with the ULTRASPEC 
fast camera, as a  new facility capable
of recording high quality LO light curves in a
geographical area where this technique was
so far precluded. We then present the first LO results
obtained at the TNT, and discuss in detail two
of them: $\alpha$~Cnc, a claimed binary star for
which we can put stringent upper limits on the
separation and flux of the putative companion; and
HR~7072, a known binary star for which
the numerous measurements available until now
seem to be not always in agreement and for which
we provide a new accurate flux and separation
determination.

\section{A New Facility for High-Time Resolution Astronomy}\label{htra}
The Thai National Observatory, which includes also the flagship
2.4\,m Thai National Telescope (TNT), is located on one of the highest
ridges of Doi Inthanon, the tallest peak in Thailand. 
At 2457\,m elevation, the site has observing conditions which compare
favorably with those of most other locations in the region in
terms of seeing and photometric conditions. It has
a dry season which runs approximately from November to April,
while the rest of the year 
is largely lost to observations
 due to high humidity and rainy conditions.
The TNT was erected in 2012 and inaugurated in January 2013.
It is a Ritchey-Chr{\`e}tien with two Nasmyth foci,  one
of them being equipped with a multi-instrument port which mounts
permanently also the ULTRASPEC instrument. 
Although a similarly named instrument based on the 
same detector has been used before,
e.g. at the ESO New Technology Telescope in Chile 
\citep{2008AIPC..984..132D, 2008SPIE.7021E..10I}, the one
described here is novel in several aspects: in particular the
optics have been specifically developed for TNT,
and  specific data acquisition modes have been developed.
ULTRASPEC at the TNT had its first light in November 2013
and is described in detail elsewhere \citep{dhillon2014}.
Here, we summarize only
its capabilities for LO work.

In order to increase the time sampling,
ULTRASPEC offers the possibility to read out only parts of the
detector. One of its
features is the 2k$\times$1k EMCCD chip, half of which is
used for imaging and half for fast frame transfer.
One or several
subarray windows can be defined and read out simultaneously,
and this is optimal to reach rates up to  about 10~Hz. For LO
measurements, however, faster rates are needed. For this,
the so-called drift-mode of the related instrument ULTRACAM
\citet{2007MNRAS.378..825D} has been specifically
adapted.
In summary, two small windows can be
selected along the same detector rows. Although their
position is freely selectable, the most efficient configuration
is to place them close to the lower edge of the detector.
A mask is then remotely positioned to cover the rest of
the detector area. As the selected windows are pushed
to the frame transfer area, the rows immediately above are shifted
down and are ready for the next integration. Other features
that we do not describe in detail here are the possibility
of on-chip rebinning and three levels of detector read-out
speed. We have successfully tested the drift mode up to rates of 450\,Hz.
 Thanks to a large memory and to
 fast fiber links between the instrument electronics,
the control computer and the data reduction machine,
it is possible to obtain uninterrupted data sequences of
any desired duration and to examine the data almost
in real time. Finally, we mention that a dedicated GPS
is used to stamp each frame with its mid-exposure time.

\section{Observations and Data Analysis}\label{obs_data}

\begin{deluxetable}{rrclclcccccrl}
\tabletypesize{\small}
\tablecaption{Summary of observed events\label{table:occlist}}

\tablehead{
\colhead{Date}&
\colhead{Time}&
\colhead{Type}&
\colhead{Source}&
\colhead{$V$}&\colhead{Sp}&
\colhead{Filter}&\colhead{Sub}&
\colhead{Bin}&\colhead{$\tau$}&\colhead{$\Delta$T}&
\colhead{S/N}&
\colhead{Notes}\\
\multicolumn{2}{c}{(UT)}   & & 
   			 & \multicolumn{1}{c}{(mag)} & 
   			 &  & 
   			\multicolumn{2}{c}{(pixels)} & 
   			\multicolumn{2}{c}{(ms)} & 
   			 & 
}
\startdata
11-Jan-14 & 17:12 & D & SAO 93721 & 5.9 & %5.2 & 
F4V & $z^\prime$ &  16x16 & 2x2 & 6.6 & 6.8 & 34 & Unresolved \\
11-Feb-14 & 16:10 & D & SAO 96543 & 7.5 & %6.1 & 
F5V & H$\alpha_{\rm B}$ &  16x16 & no & 11.8 & 12.1 & 10 & Unresolved \\
12-Mar-14 & 18:26 & D & SAO 97913 & 6.3 & %4.5 & 
K0III & H$\alpha_{\rm B}$ &  16x16 & 2x2 & 6.6 & 6.8 & 45 & Unresolved \\
09-Apr-14 & 13:20 & D & $\alpha$~Cnc & 4.3 & %4.1\footnote{I band} & 
A5m & H$\alpha_{\rm N}$ & 16x16 & 2x2 & 6.6 & 6.8 & 104 & Not Binary \\
20-Apr-14 & 21:44 & R & HR~7072 & 6.5 & %4.6 & 
A1V+K1III & H$\alpha_{\rm B}$ &  32x32 & 2x2 & 12.3 & 12.8 & 18 & Binary \\
\enddata
\end{deluxetable}

The first LO observations recorded from TNT with ULTRASPEC
are listed in Table~\ref{table:occlist},
in chronological order.
D and R refer to disappearances and reappearances, respectively.
 The magnitudes and
spectra are quoted from Simbad. Concerning the filters,
we adopted mainly narrow-band ones in the attempt to 
reduce the lunar background. The latter
is strongly wavelength dependent, and also
diffraction is inherently chromatic. Thus, the red part
of the spectrum is usually better suited for LO
observations.
In the table,
H$\alpha_{\rm N}$ and H$\alpha_{\rm B}$ differ slightly in 
central wavelength (6564 and 6554 \AA, respectively),
and in FWHM (54 and 94 \AA, respectively). 
We also employed a standard SDSS $z^\prime$ filter.
The columns Sub and Bin list, respectively,
the size of the detector subarray
adopted in the drift-mode and the on-chip
rebinning - so that Sub 32x32 and Bin 2x2 would 
effectively result in a 16x16 output.
The frame integration time and the sampling time 
between frames are
denoted by 
$\tau$  and $\Delta$T in the table.
This latter is the average value across all
frames. In reality, there are small
variations
(always $< 1$\%) in the time differences between
subsequent frames.
In our data analysis, the actual 
individual time stamps are used.
S/N is the signal-to-noise ratio, measured as the unocculted stellar
signal divided by the rms of the fit residuals.

The raw binary output from ULTRASPEC is then
converted to FITS cubes, with the first two dimensions
set by the combination of Sub and Bin, and thousands of
frames long. From this point,
our data analysis procedure follows closely the
methods already described in our previous papers,
see e.g. \citet{2014AJ....147...57R} and references
therein. 
In summary, we adopt a mask extraction that
allows us to reject unnecessary signal (and related noise)
from the background, and end up with
fast photometry sequences which are further restricted
to very few seconds around the  events.

We use a model-independent maximum-likelihood
\citep[CAL,][]{1989A&A...226..366R}
method to estimate the brightness profile of the
source, for example to detect possible multiple
components in the light curve.
A model-dependent least-squares method, whose
convergence in $\chi^2$ is driven by noise components 
derived
from the unocculted and totally occulted portions of
the light curve, is used to derive precise
values and errors of parameters such as intensities
and separation in a binary model
\citep{richichi96}. This reference also describes 
additional features used in our
fits such as the modelling of low-frequency scintillation.

\section{Results}
The first three sources in Table~\ref{table:occlist}
were found to be point-like, with upper limits on the 
angular size of 2.4 and 1.5\,mas for
SAO~93721 and SAO~97913, respectively.
These are consistent with the values expected from their
spectral types and distances. The S/N of the
SAO~96543 light curve was insufficient for an upper limit determination.
There were no previous
literature reports on possible binarity for two of these sources, while
SAO~93721 is listed in the Washington Double Star Catalog. It has
a faint, distant companion at about 170$\arcsec$ which is not
included in our observation, and is also listed as a spectroscopic
binary which we could not resolve.
In the following, we concentrate on the remaining two stars
in our list.

Although these first measurements are few, Table~\ref{table:occlist}
can be used for some initial estimate of the performance of
LO with ULTRASPEC at the TNT. Using S/N=1 as a detection limit,
the sensitivity achieved in the 5 light curves ranged from 9.3 to 10.4\,mag,
with a dynamic range that in the best case was 5\,mag. The ability to
resolve close companions was tested by means of simulations on the data
of SAO~93721 and $\alpha$~Cnc. We conclude that hypothetical companions
with 1:1 flux ratio could have been detected as close as 2~mas.

\subsection{Alpha Cancri}\label{alfcnc}
\begin{figure}
%\vspace{302pt}
 \includegraphics[angle=-90, width=84mm]{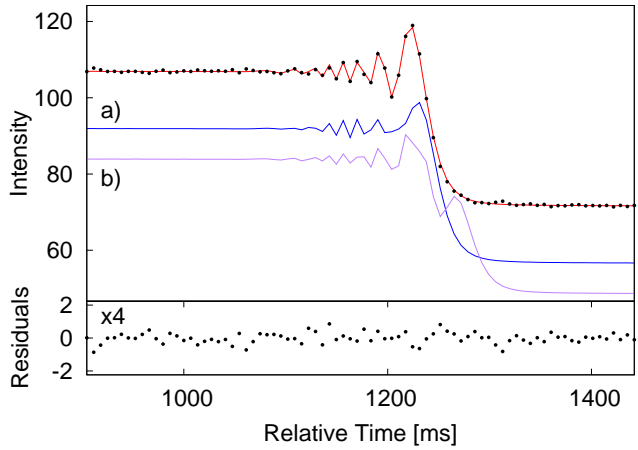}
 \caption{Top panel: occultation data (dots) for 
 $\alpha$~Cnc, and best
 fit by a point-like source (solid line). 
  The fit residuals
 are shown in the lower panel, enlarged for clarity.
 The curves
 labelled a) and b) are models for a binary star
 with equal components and projected separations of 3 and
 10\,mas, respectively. They are shown, shifted by arbitrary vertical
 offsets for clarity, to prove
 the inconsistency of the data with such scenarios.
 A color version of this figure is available online.
}\label{fig:alfcnc}
\end{figure}

We obtained a good quality (S/N=104) light curve of the
bright star $\alpha$~Cnc (HR~3572, Acubens), shown in
Fig.~\ref{fig:alfcnc}, along the position angle (PA) 77$\degr$. 
The diffraction fringe pattern is well resolved thanks to
the use of a narrow band filter and also to a contact
angle of 40$\degr$ that contributed to slow down the
apparent fringe motion. We could thus 
accurately measure the local limb slope, found to be close to 0$\degr$,
obtaining in turn a reliable conversion from time to angular scale.
The star is known to have a  companion
forming the ADS~7115 pair; this is however about 10$\arcsec$
away and several magnitudes fainter, and we do not concern
ourselves with it here.

More interestingly, $\alpha$~Cnc has been claimed as a close
double, mainly from previous LO observations but also from
the analysis of Hipparcos data. 
These are listed in the Fourth  Catalog of 
Interferometric Measurements of Binary Stars
(by Hartkopf, Mason and Wycoff, available online\footnote{http://www.usno.navy.mil/USNO/astrometry/optical-IR-prod/wds/int4};
INT4 hereafter). The two claims from previous occultations come
from observations by amateur astronomers with small
telescopes, and are not well documented. They claim
a binary with equal components separated by 50\,mas
in one case (with no PA listed), or 4.2\,mas along
PA=113$\degr$ in the other. We note that the ability
to measure such a small separation by LO usually demands
a rigorous treatment of high quality data.
The Bright Star Catalogue mentions in a note
that $\alpha$~Cnc is a LO binary with $0\farcs1$ separation, but
without a reference.
Several speckle measurements with the SOAR 4.1\,m telescope
\citep{{2010AJ....139..743T},{2012AJ....143...42H}}
have not detected the companion, with upper limits
$\le 0\farcs15$ on the  angular separation
and 4 to 6\,mag on the flux difference. In fact,
for $\Delta$m$\le 1$\,mag the separation limit is
25\,mas (Tokovinin, priv. comm.).

Our data appear to be the first LO light curve recorded
of this bright star with professional equipment at
a medium-sized telescope. We can reliably exclude two components 
of similar brightness, to projected separations as small as
3\,mas (see Fig.~\ref{fig:alfcnc}). In fact, an analysis
by a method developed for unresolved 
sources \citep[see][]{richichi96,apjssur} results in a
limit of 1.35\,mas on the angular size of the star.
This is  consistent with a stellar dimension roughly
comparable to solar and a distance of 53\,pc as derived by Hipparcos.
For a simple comparison, we assume that the 50\,mas 
binary separation
previously claimed was at a random position angle,
and we consider an average projection factor $2/ \pi$.
The resulting estimated 79\,mas true separation, against our
upper limit of 3\,mas, would indicate a likelihood
of $1-({\rm acos}(3/79)/90\degr)\approx 2$\% that
our lack of detection was due to projection effects.
We note that the previously claimed
binary separation would imply a period of very few
years, and therefore any link between the geometry
during our measurement
in 2014 and the previous ones (as much as 30 years
earlier) would not be preserved.
The S/N of our light curve can also be used to
assess an upper limit in flux on the presence of
a companion with a more significant separation, of
about 5\,mag in the red part of the spectrum.
We mention also two light curves $\alpha$~Cnc, the disappearance
and reappearance of the same event, observed with
a 30-inch telescope in a 5214\,{\AA}  interference filter
by \citet{1978AJ.....83.1100A}. The authors also found
no evidence of binarity, but given the small telescope size
scintillation was significant and the constraint on
a putative companion was weaker than that from our data.

The other indication of binarity in $\alpha$~Cnc comes
from Hipparcos, which detected an acceleration
in the proper motion, interpreted
as due to a companion. However no orbital solution
could be derived and the nature of the companion (separation and mass,
hence brightness) remains undetermined,
and likely unrelated to the putative companion
claimed by previous LO reports. 
\citet{2007AA...464..377F}, after examining several
radial velocities catalogs, found no evidence
of $\alpha$~Cnc being a spectroscopic binary.

\subsection{HR 7072}\label{hr7072}
\begin{figure}
%  \vspace*{174pt}
  \includegraphics[angle=-90, width=84mm]{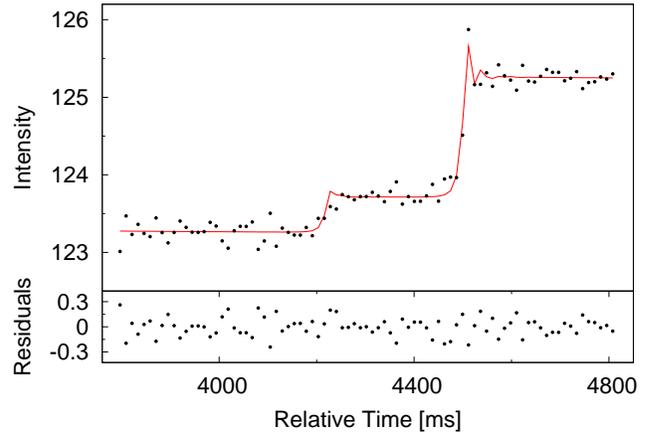}
  \caption{The light curve data for HR~7072 (dots) and the best
  fit by a binary star model (solid line). The fit residuals
 are shown in the lower panel.
  }\label{fig:hr7072-lo}
\end{figure}

Our reappearance light curve of HR~7072 (Kui 88AB, HIP 92301) is 
shown in  Fig.~\ref{fig:hr7072-lo} and it clearly reveals
a companion. We find a flux ratio of 
3.38$\pm$0.01, and a 
projected separation of 90.7$\pm$0.2\,mas along
PA=238$\degr$. In order to successfully record
the reappearance event, we opted for a larger
subarray which resulted in a 
relatively slow sampling. Thus, 
we are not able to determine 
independently the speed of the fringe pattern,
and in turn we have an uncertainty on the 
the local limb slope and correspondingly on 
the PA and projected separation. However, an analysis
of the lunar limb at the given point of contact
and libration angles was made using the
altimetry data from the Kaguya probe (Herald,
priv. comm.), and no significant local limb slope
was found. 

HR~7072 is a known sub-arcsecond binary,
with over 30 previous astrometric and LO determinations, 
starting from the initial discovery by
\citet{1950AJ.....55..153W} and extending over 54 years.
A  list can be found
in INT4, and additionally we mention \citet{1958JO.....41..109M}.
Our result extends the time coverage to almost 65 years.
An orbital solution has not yet been established, 
and in fact the interpretation of the available data is not 
straightforward. 

\begin{figure*}
%  \vspace*{174pt}
  \includegraphics[angle=-90, width=150mm]{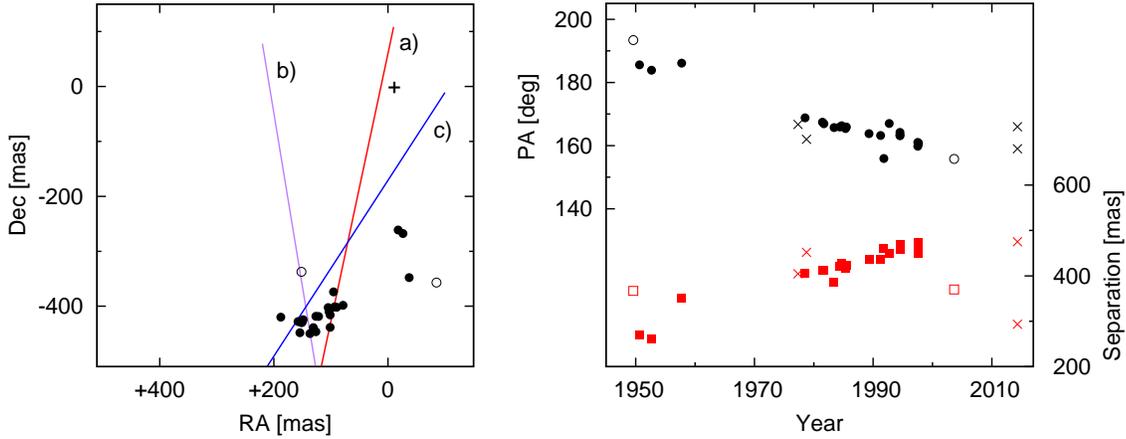}
  \caption{
  Left: astrometric (circles) and LO (lines) determinations
  of the position of the companion of HR~7072 with respect
  to the primary (cross). The LO are a) \citet{1978AJ.....83.1100A},
  b) \citet{1980AJ.....85..478E},
  c) our measurement.
  Right: the same data, converted 
  to PA (top) and separation (bottom) and plotted as a function of time.
  In this plot, the crossed symbols represent estimated
  LO positions, extrapolated from the lines in the left panel.
  The open symbols in both plots are discussed in the text.
  Errors are mostly not available.
   A color version of this figure is available online.
  }\label{fig:hr7072-orb}
\end{figure*}

The left panel of Fig.~\ref{fig:hr7072-orb} shows the RA, Dec position of the companion.
For the LO results,
which only provide a projected separation, we have plotted the lines
that represent the loci of equivalent position. At a first glance,
it can be noticed that the positions tend to cluster around the
[+150,$-$400]\,mas region, with the notable exception of the points 
on the negative RA half-plane which are all prior to 1960.
These were visual determinations, while later measurements
were obtained by speckle interferometry,
The right panel of Fig.~\ref{fig:hr7072-orb} puts the measurements,
expressed now as PA and separation,
along a time sequence. It can be seen that both quantities
follow an almost linear evolution with time, however with the
noticeable exception of two points. These are marked as
outlined, rather than solid, symbols, on both panels of Fig.~\ref{fig:hr7072-orb}.
The first one is the discovery measurement by \citet{1950AJ.....55..153W},
the second one is the latest measurement, prior to ours, obtained
by \citet{2004AJ....128.3012M}. The former was obtained by
visual interferometry, the latter by speckle, but both at relatively
smaller telescopes than the majority of the other determinations.
We remark that most of the literature values do not have
an associated error, so that it is difficult to verify
quantitatively possible discrepancies. 

In the right panel of  Fig.~\ref{fig:hr7072-orb}
we have made an attempt to estimate PA and separation
for the two previous LO measurements and our own, based
also on a visual fit to the rest of the points in the left panel.
It can be seen that in this view the
apparent discrepancy of the \citet{1950AJ.....55..153W} measurement
could in fact be reconciled given a reasonable error, as expected
from the method. The \citet{2004AJ....128.3012M} measurement
is critical, in that 
taken at face value it points to a significant turn in the
orbit. Accordingly, we provide two PA-sep equivalents for
our LO result: one fits a linear trend in time of PA and separation,
consistent with a very long orbital period. The other follows
the indication of the \citet{2004AJ....128.3012M} measurement, 
and reinforces the view of a closed orbit which would then seem to
have a period of few 10$^2$~y. 

\begin{figure}
%  \vspace*{174pt}
  \includegraphics[angle=-90, width=84mm]{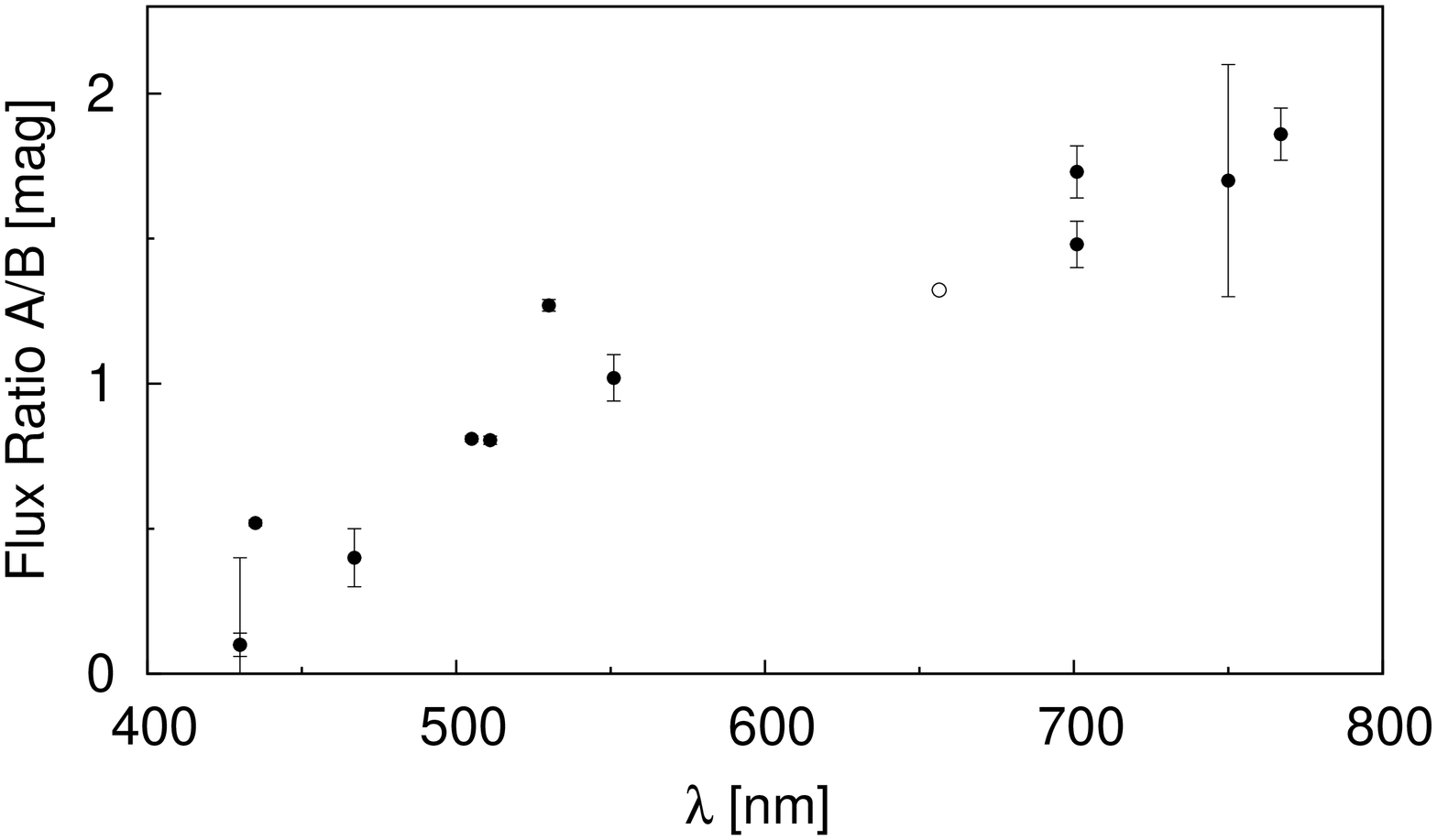}
  \caption{The flux ratio of the two components of HR~7072,
  as a function of wavelength. Solid dots are previous results,
  the outlined circle is our measurement. Some errorbars are
  too small to be seen clearly on this scale.
  }\label{fig:hr7072-flux}
\end{figure}

Concerning the flux ratio, the available data span
the range from 400 to 800\,nm, and are quite
consistent in their trend indicating that significant
variability of either component can probably be excluded,
see Fig.~\ref{fig:hr7072-flux}.
Our own
determination is in close agreement with this trend.
The conclusion is that one of the two stars is significantly
bluer than the other.
\citet{1985A&AS...59...15A} measured the radial velocity
of this system on two dates only, separated by just one year.
They did not detect any variations. They quote the spectrum
as composed of A1 and K0. \citet{1988mcts.book.....H}
reported A1V and K1III.
 This, together with the
integrated color of $V-K=2.8$\,mag,
points to the fact that the two stars have similar brightness
in the blue, but that at longer wavelengths the K0 component
is the primary. This is consistent with \citet{1991A&A...252..229J},
who found a 12\,$\mu$m excess for this star and mentioned
binarity as a possible explanation.

\section{Conclusions}
We reported the first LO observations from
the Thai National Telescope equipped
with the ULTRASPEC instrument. Using the specifically
developed drift-mode,
time sampling of few milliseconds can be achieved. 
In the first observing season of this new facility,
we have recorded 5 LO light curves and we have
discussed in detail two of them. 

In the case of $\alpha$~Cnc a
duplicity has been previously claimed, although
not convincingly demonstrated. We do not
detect the companion, with an upper limit of
3\,mas in projected separation and 5\,mag in flux ratio.
In the case of {HR~7072}, our measurement goes
to complement a set of over 30 previous determinations, extending
the time coverage from 54 to 65 years. We
show that the data are still ambiguous, pointing to
the detection of
an orbital arch and
a few 10$^2$~y period, or alternatively to 
a much wider and longer orbit. Observations by adaptive optics would
quickly confirm one of the two scenarios.
The flux ratios are consistent with an early A and
an early K spectral types, with similar brightness
in the blue but with the K type component dominating
at longer wavelengths.

The TNT with ULTRASPEC in drift mode has successfully demonstrated
high performance LO
results, with a sensitivity estimated
at $I \approx 10$\,mag at S/N=1, an angular resolution
close to 1\,mas, and a dynamic range of at least
5\,mag. This facility is especially attractive
in consideration of the fact the LO events are
observable only along restricted ground tracks, and that
no comparable capabilities existed until now
in South-East Asia.

\acknowledgments
VSD and TRM acknowledge the support of the Royal Society and the
Leverhulme Trust for the operation of ULTRASPEC at the TNT.
We are grateful to Dr. A.~Tokovinin for useful discussions.
The estimation of the lunar limb slope for the occultation
of HR~7072 was provided by Dr. D.~Herald.
This research made use of the Simbad database,
operated at the CDS, Strasbourg, France.

\clearpage

\end{document}